\documentclass{iaus}
\usepackage{graphicx}

  \checkfont{eurm10}
  \iffontfound
    \IfFileExists{upmath.sty}
      {\typeout{^^JFound AMS Euler Roman fonts on the system,
                   using the 'upmath' package.^^J}%
       \usepackage{upmath}}
      {\typeout{^^JFound AMS Euler Roman fonts on the system, but you
                   dont seem to have the}%
       \typeout{'upmath' package installed. iaus.cls can take advantage
                 of these fonts,^^Jif you use 'upmath' package.^^J}%
      }
  \else
  \fi

  \checkfont{msam10}
  \iffontfound
    \IfFileExists{amssymb.sty}
      {\typeout{^^JFound AMS Symbol fonts on the system, using the
                'amssymb' package.^^J}%
       \usepackage{amssymb}%
         
       \let\ge=\geqslant  
      }{}
  \fi

  \IfFileExists{amsbsy.sty}
    {\typeout{^^JFound the 'amsbsy' package on the system, using it.^^J}%
     \usepackage{amsbsy}}
    {}

\newsavebox{\astrutbox}
\sbox{\astrutbox}{\rule[-5pt]{0pt}{20pt}}

\title[A Hierarchy of Voids]
      {A Hierarchy of Voids}

\author[R. van de Weygaert, R. Sheth \& E. Platen]%
{Rien van de Weygaert$^1$, Ravi Sheth$^{2,3}$ \and Erwin Platen$^1$}

\affiliation{$^1$ Kapteyn Instituut, Rijksuniversiteit Groningen, the Netherlands\\
$^2$ Dept. of Physics and Astronomy, University of Pittsburgh,U.S.A.\\
$^3$ NASA/Fermilab Astrophysics Group, Batavia, IL 60510-0500, U.S.A.}

\pubyear{2004}
\volume{195}
\pagerange{1--8}
\date{?? and in revised form ??}
\jname{Outskirts of Galaxy Clusters: intense life in the suburbs}
\editors{A. Diaferio, ed.}
\begin{document}

\maketitle

\begin{abstract}
We present a model for the distribution of void sizes and its 
evolution within the context of hierarchical scenarios of 
gravitational structure formation. For a proper description of the 
hierarchical buildup of the system of voids in the matter distribution, 
not only the {\it void-in-void} problem should be taken into account, 
but also that of the {\it void-in-cloud} issue. Within the context 
of the excursion set formulation of an evolving void hierarchy 
is one involving a {\it two-barrier} excursion problem, unlike the 
{\it one-barrier} problem for the dark halo evolution. This leads 
to voids having a peaked size distribution at any cosmic epoch, centered on 
a characteristic void size that evolves self-similarly in time, 
in distinct contrast to the distribution of virialized halo masses in not 
having a small-scale cut-off. 
\end{abstract}

\section{Introduction: Excursions}

\noindent Hierarchical scenarios of structure formation have been very succesfull 
in understanding the formation histories of gravitationally bound 
virialized haloes. Particularly compelling has been the formulation of a formalism 
in which the collapse and virialization of overdense dark matter halos within the 
context of hierarchical clustering can be treated on a fully analytical basis. This 
approach was originally proposed by Press \& Schechter (1974), which 
found a particularly useful and versatile formulation and modification in the 
the {\it excursion set formalism} (Bond et al.~1991). 

It is based on the assumption that for a structure to reach a particular nonlinear 
evolutionary stage, such as complete gravitational collapse, the sole condition 
is that its {\em linearly extrapolated primordial density} should attain a certain 
value. The canonic example is that of a spherical tophat overdensity collapsing once 
it reaches the collapse barrier $\delta_c\approx 1.69$. The successive contributions 
to the local density by perturbations on a (mass) resolution scale $S_m$ may 
be represented in terms of a density perturbation random walk, the 
cumulative of all density fluctuations at a resolution scale smaller 
than $S_m$. By identifying the largest scale at which the density passes 
through the barrier $\delta_c$ it is possible (1) to infer at any cosmic epoch the 
mass spectrum of collapsed halos and (2) to reconstruct the merging history of 
each halo (see Fig.~3, lefthand). 
\begin{figure}
\vskip -3.0truecm
\mbox{\hskip -0.5truecm\includegraphics[width=14.2cm]{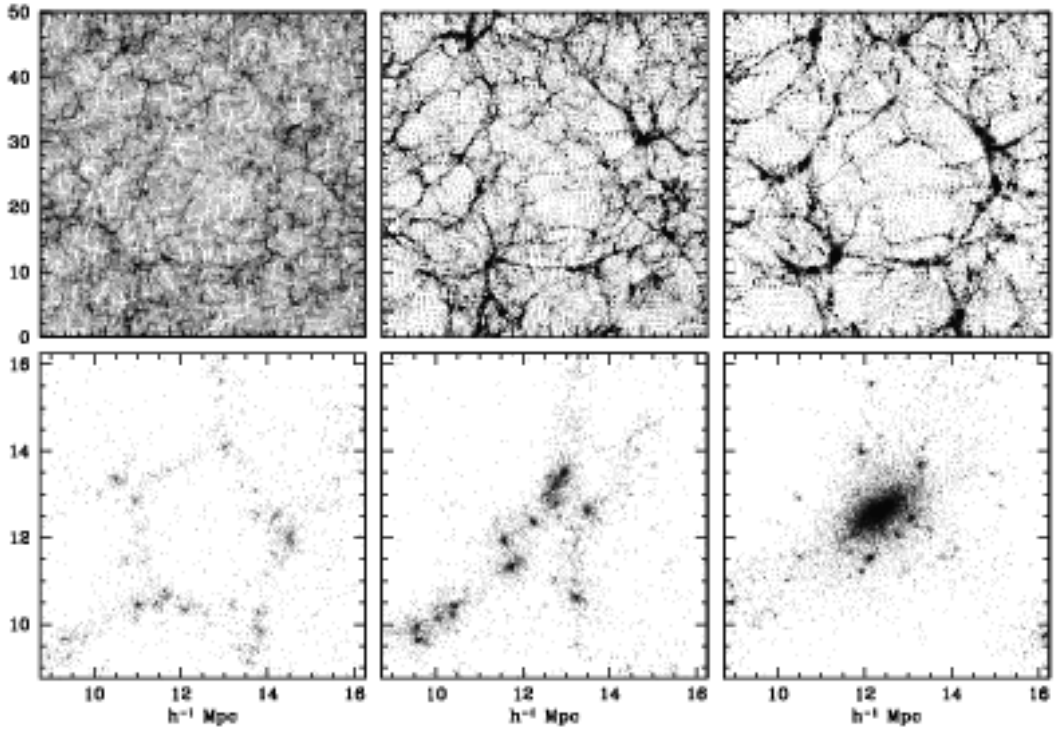}}
\vskip -1.5truecm
\caption{Illustration of the two essential ``void hierarchy modes'': (top) the 
{\em void-in-void} process (top), with a void growing through the 
merging of two or more subvoids; (bottom) the {\em void-in-cloud} process:  
a void demolished through the gravitational collapse of embedding region.}
\vskip -0.2truecm
\label{voidproc}
\end{figure}
In this study we demonstrate that also the formation and evolution of foamlike patterns 
as a result of the gravitational growth of primordial density perturbations is liable 
to a succesfull description by the excursion set analysis. A slight extension and 
elaboration on the original formulation enables us to frame an analytical 
theory explaining how the characteristic observed weblike Megaparsec scale 
galaxy distribution, and the equivalent frothy spatial matter distribution seen 
to form in computer simulations of cosmic structure formation, are natural 
products of a hierarchical process of gravitational clustering. This is 
accomplished by resorting to a complementary view of clustering evolution in 
which we focus on the evolution of the {\em voids} in the Megaparsec galaxy and 
matter distribution, spatially {\it th\'e} dominant component (see e.g. Muller \& 
Maulbatsch, these proceedings). 
\begin{figure}
\vskip 0.5cm
\mbox{\hskip 0.8truecm\includegraphics[width=12.0cm]{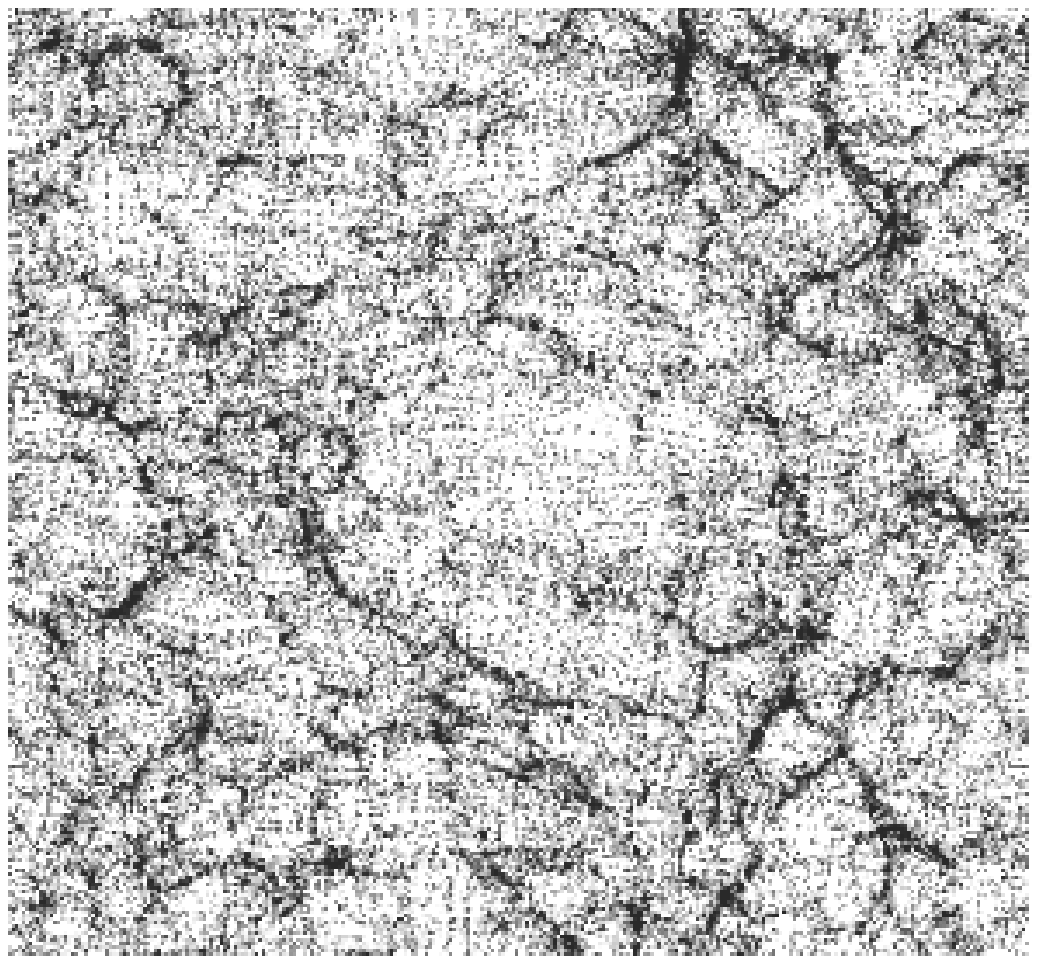}}
\vskip -2.0cm
\mbox{\hskip 0.0truecm\includegraphics[width=6.5cm]{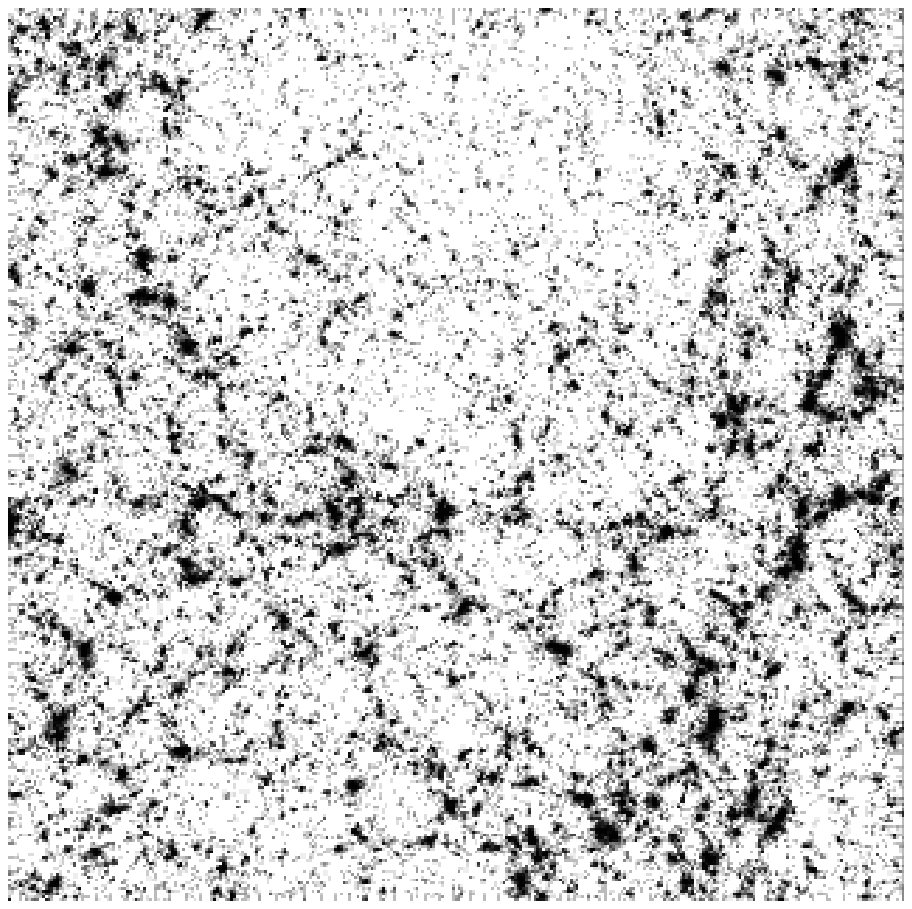}}
\vskip -6.5cm
\mbox{\hskip 7.0truecm\includegraphics[width=6.5cm]{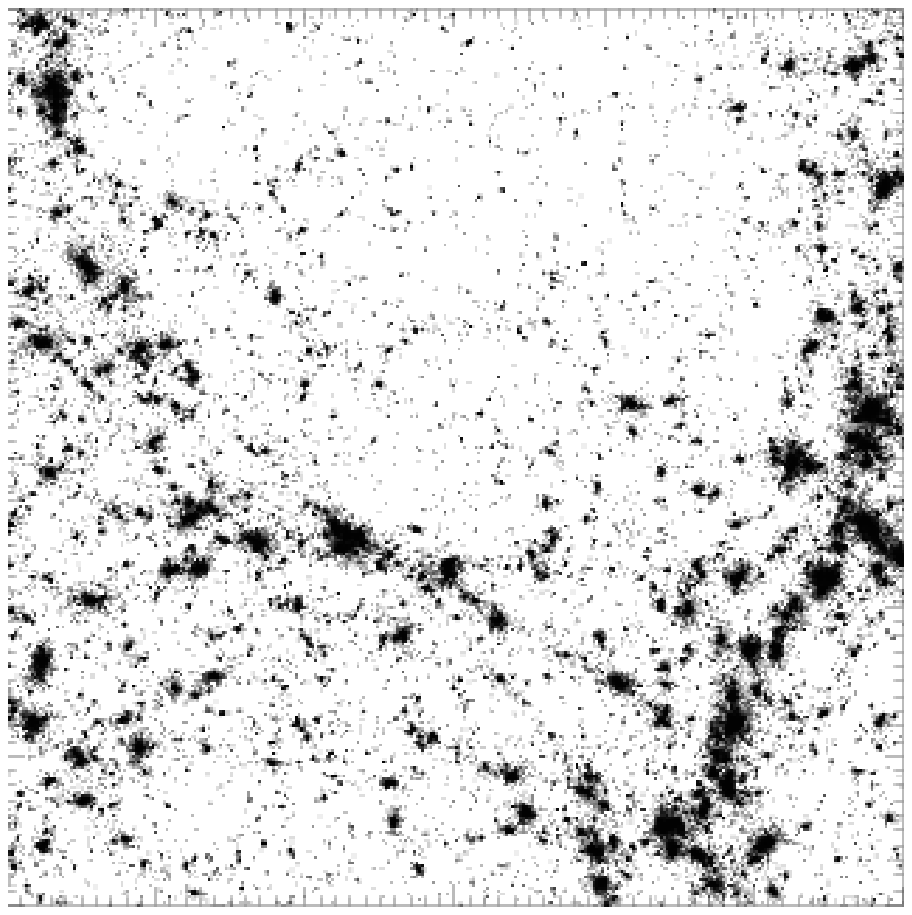}}
\vskip 0.0cm
\caption{Identification of void collapse sites: near the boundary of an expanding 
voids small voids get squashed and sheared. Zoom-in on central lower region of 
large void (top image) at 2 different timesteps, showing the void 
compression process.}
\vskip -1.0truecm
\end{figure}

\section{Void Evolution: the two processes}

\noindent Primordial underdensities are the progenitors of voids. Because underdensities 
are regions of suppressed gravitational attraction, representing an effective repulsive 
gravity, matter flows out of their interior and moves outward to the edges of these 
expanding {\em voids}. Voids {\em expand}, become increasingly {\em empty} and 
develop an increasingly {\em spherical} shape (Icke~1984). Matter 
from the void's interior piles up near the edge: usually a ridge forms 
around the void's rim and at a characteristic moment the void's interior 
shells take over the outer ones. At this {\em shell-crossing} epoch the void 
reaches maturity and becomes a nonlinear object expanding self-similarly, the 
implication being that the majority of observed voids is at or near this stage  
(Blumenthal et al.~1992). As voids develop from underdensities 
in the primordial cosmic density field, the interaction with internal substructure and 
external surrounding structures translates into a continuing process of 
hierarchical void evolution (Dubinski et al.~1993, Van de Weygaert \& 
Van Kampen~1993,Colberg et al. 2004). 

\begin{figure}
\vskip -0.truecm
\mbox{\hskip -0.5truecm\includegraphics[width=16.0cm]{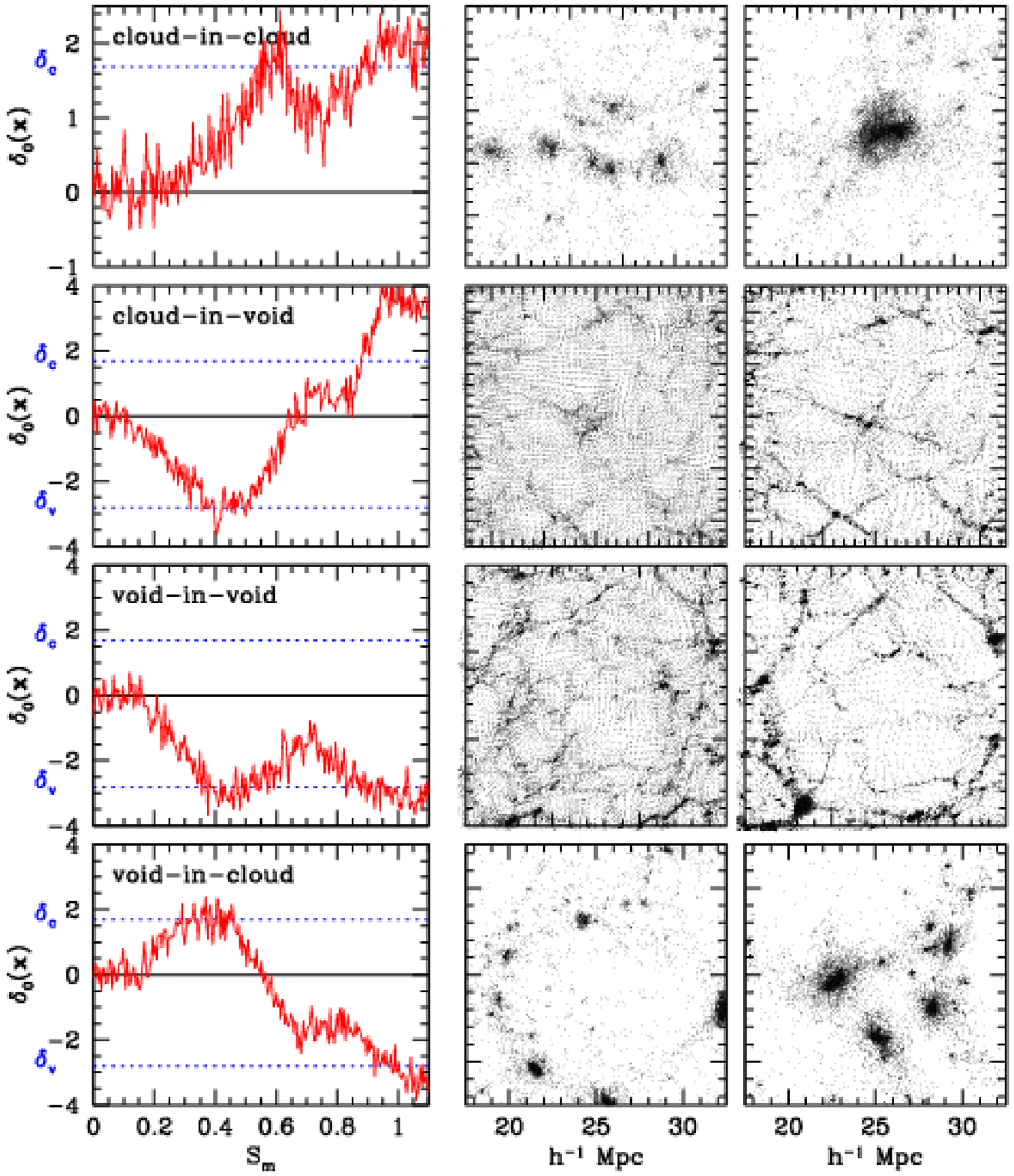}}
\vskip -0.cm
\caption{Four mode (extended) excursion set formalism. 
Each row illustrates one of the four basic modes of hierarchical 
clustering: the {\em cloud-in-cloud} process, 
{\em cloud-in-void} process, {\em void-in-void} process and 
{\em void-in-cloud} process (from top to bottom). 
Each mode is illustrated using three frames. Leftmost panels show `random walks': the local density 
perturbation $\delta_0({\bf x})$ as a function of (mass) 
resolution scale $S_m$ at an early time in an N-body simulation of cosmic structure formation. 
In each graph, the dashed horizontal lines indicate the 
{\em collapse barrier} $\delta_{\rm c}$ and the shell-crossing 
{\em void barrier} $\delta_{\rm v}$. 
The two frames on the right show how the associated particle 
distribution evolves. Whereas halos within voids may be observable (second row depicts a 
halo within a larger void), voids within collapsed halos are not 
(last row depicts a small void which will be squeezed to small size 
as the surrounding halo collapses). It is this fact which makes the 
calculation of void sizes qualitatively different from that usually 
used to estimate the mass function of collapsed halos.}
\vskip -0.2truecm
\end{figure}

The evolution of voids resembles that of dark halos in that large 
voids form from mergers of smaller voids that have matured at an 
earlier cosmic time (Fig.~1, top row). However, in contrast to dark halos, 
the fate of voids is ruled by two processes. Crucial is the realization that the evolving 
void hierarchy does not only involve the {\em void-in-void} process but 
also an additional aspect, the {\em void-in-cloud} process.
Small voids may not only merge into larger encompassing underdensities, 
they may also disappear through collapse when embedded within a larger scale 
overdensity (Fig.~1, bottom row). In terms of the {\em excursion set approach}, 
it means that the {\em one-barrier} problem for the halo population has to be extended 
to a more complex {\em two-barrier problem}. Voids not only should ascertain 
themselves of having decreased their density below the {\em void barrier} $\delta_v$ of 
the {\em shell-crossing} transition. For their survival and sheer existence it is 
crucial that they take into account whether they are not situated within a collapsing 
overdensity on a larger scale which crossed the collapse barrier $\delta_c$. They 
should follow a random walk path like type ``3'' in Fig.~3, rather than 
the {\em void-in-cloud} path ``4''). The repercussions of this are far-reaching and 
it leads to a major modification of the void properties and distribution.

\section{Void Distribution: Universal, Peaked and Self-Similar}

\noindent Analytically, the resulting expression follows by evaluating the 
fraction of walks which first cross $\delta_{\rm v}$ at $S$, and which do not 
cross $\delta_{\rm c}$ until after they have crossed $\delta_{\rm v}$ 
(see Fig.~3, 3rd \& 4th row). This distribution 
${\cal F}(S,\delta_{\rm v},\delta_{\rm c})$ is given by 
(Sheth \& van de Weygaert~2003), 
\begin{equation}
 S\,{\cal F}(S,\delta_{\rm v},\delta_{\rm c}) = 
           \sum_{j=1}^\infty {j^2\pi^2 D^2\over\delta_{\rm v}^2/S}\,
           {\sin(j\pi D)\over j\pi} \times \exp\left(-{j^2 \pi^2 D^2\over 2\,\delta_{\rm v}^2/S}\right),
 \label{vfvoid}
\end{equation}
in which the relative impact of void and halo evolution on the hierarchically evolving 
population of voids is parameterized through $D\equiv (|\delta_{\rm v}|/(\delta_{\rm c}-\delta_{\rm v})$. A smaller value of $D$ corresponds to a diminished importance of the 
{\em void-in-cloud} process. The above expression may be approximated by the more 
accessible expression, for values of $\delta_{\rm c}/|\delta_{\rm v}|\ge 1/4$, using 
the ``self-similar'' parameter $\nu\,\equiv\,\delta_{\rm v}^2/S\,$ 
\begin{equation}
 \nu f(\nu) \approx \sqrt{\nu\over 2\pi}\,
 \exp\left(-{\nu\over 2}\right)\,
 \exp\left(-{|\delta_{\rm v}|\over\delta_{\rm c}}\,
 {D^2\over 4\nu}-2{D^4\over\nu^2} \right).
 \label{vfvapprox}
\end{equation}

\begin{figure}
\vskip -1.75truecm
\centering\mbox{\hskip -0.8truecm\includegraphics[width=14.2cm]{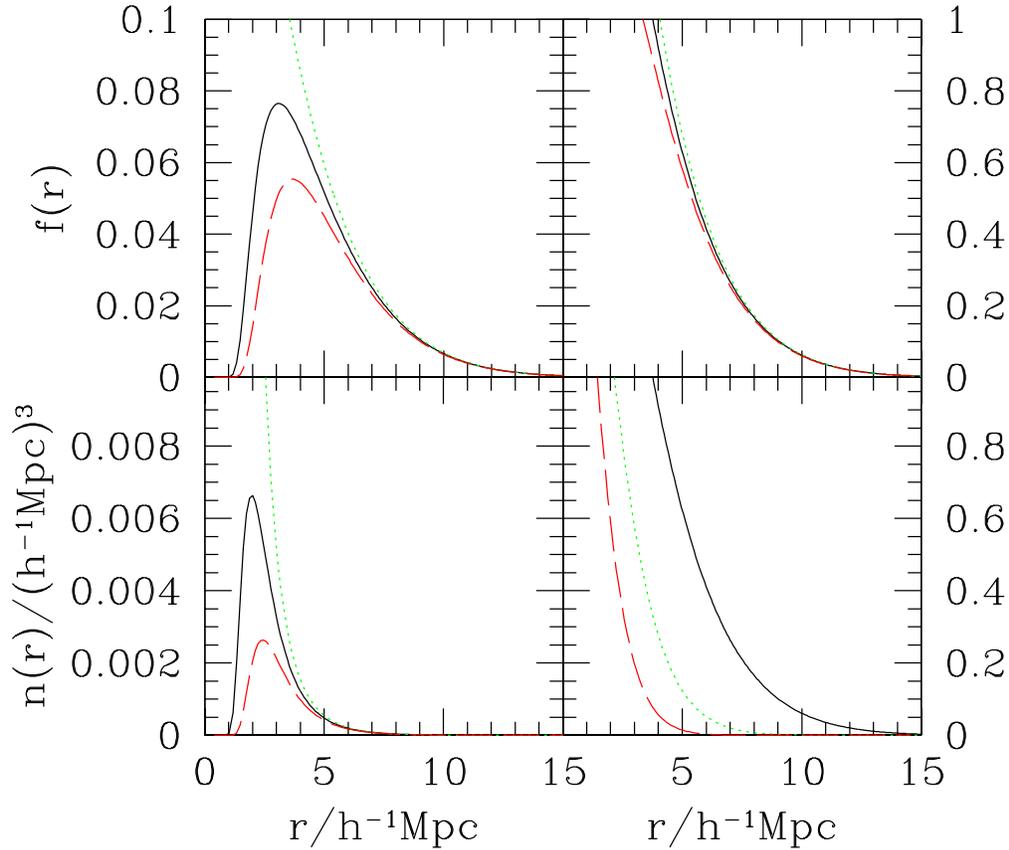}}
\vskip -1.1truecm
 \caption{\small{\it Distribution of predicted void radii predicted by, in an 
 Einstein de-Sitter model with 
 $P(k)\propto k^{-1.5}$, normalized to $\sigma_8=0.9$ at $z=0$.  
 Top left panel shows the mass fraction in voids of radius $r$.  
 Bottom left panel shows the number density of voids of radius $r$.  
 Note that the void-size distribution is well peaked about a 
 characteristic size provided one accounts for the void-in-cloud 
 process.  Top right panel shows the cumulative distribution of the 
 void volume fraction. 
 Dashed and solid curves in the top panels and bottom left panel 
 show the two natural choices for the importance of the void-in-cloud 
 process discussed in the text:  $\delta_{\rm c}=1.06$ and 1.686, 
 with $\delta_{\rm v}=-2.81$.  Dotted curve shows the result of 
 ignoring the {\em void-in-cloud} process entirely.
 Clearly, the number of small voids decreases as the ratio of 
 $\delta_{\rm c}/|\delta_{\rm v}|$ decreases. 
 Bottom right panel shows the evolution of the cumulative void volume 
 fraction distribution. The three curves in this panel are for 
 $\delta_{\rm c}=1.686(1+z)$, where $z=0$ (solid), 0.5 (dotted) 
 and~1 (dashed).}}
\vskip -0.25truecm
\end{figure}
\noindent The resulting distributions, for various values of $D$, are shown in 
Figure~4. The void size distribution is clearly {\em peaked about a 
characteristic size}: the {\em void-in-cloud} mechanism is responsible for the demise of 
a sizeable population of small voids. The halo mass distribution diverges towards small 
scale masses, so that in terms of numbers the halo population is dominated by 
small mass objects. The void population, on the other hand, is ``void'' of small voids and 
has a sharp small-scale cut-off. 

Four additional major observations readily follow from this analysis: 
($\bullet$) At any one cosmic epoch we may identify a 
{\em characteristic void size} which also explains why in the present-day foamlike 
spatial galaxy distribution voids of $\sim 20-30h^{-1}\hbox{Mpc}$ are the predominant 
feature; ($\bullet$) The {\em void distribution evolves self-similarly} and the 
{\em characteristic void size increases with time}: the larger voids 
present at late times formed from mergers of smaller voids which constituted the 
dominating features at earlier epochs (Fig.~4, top panels); 
($\bullet$) Volume integration shows that at any given time {\em the population of voids 
approximately fill space}, apparently squeezing the migrating high-density matter in 
between them; ($\bullet$) As the size of the major share of voids will be in the order 
of that of the characteristic void size this observation implies that the 
{\it cosmic matter distribution} resembles a {\it foamlike packing of spherical voids of 
approximately similar size and excess expansion rate}. 

To develop this generic picture into an encompassing view of the formation and evolution of the 
foamlike galaxy distribution various additional issues need to be adressed. One issue 
concerns the translation from matter into galaxy distribution, and thus the question of 
galaxy formation within voidlike environments. In this respect it should be noted 
that the suggested scale of the peaked matter distribution still falls short of that of the 
observed galaxy distribution (see e.g. Vogeley 2004). Another issue is the 
identification and location of 
the collapsing void population. Recent results (Fig.~2) have indicated that most of 
these events take place near and on the boundary of the large expanding voids: most 
small voids appear to get squashed and sheared (see Van de Weygaert \& Babul 1996) there 
where the matter currents out of voids collide with the surrounding overdensities. 

In all, from this study an enticing image of cosmic structure evolution emanates.  
A continuously evolving hierarchy of voids produces a dynamical foamlike pattern whose 
characteristic dimension grows continuously along with the evolution of cosmic 
structure, a Universe which at any one cosmic epoch is filled with bubbles whose 
size corresponds to the scale just reaching {\em maturity}.

\end{document}